\documentclass[twocolumn,showpacs,prb,10pt]{revtex4}
\usepackage[english]{babel}
\usepackage{amsmath,amssymb}
\usepackage{times}
\usepackage{epsfig}

\begin{document} 

\title {Resistivity and optical conductivity of cuprates within the
  $t$ - $J$ model}

\author{M. M. Zemlji\v c$^{1}$ and P. Prelov\v sek$^{1,2,3}$}
\affiliation{$^1$J.\ Stefan Institute, SI-1000 Ljubljana, Slovenia}
\affiliation{$^2$ Faculty of Mathematics and Physics, University of
Ljubljana, SI-1000 Ljubljana, Slovenia} 
\affiliation{$^3$
Max-Planck-Institut f\" ur Festk\" orperforschung, D-70569, Germany}

\date{\today}
 
\begin{abstract} 
  
The optical conductivity $\sigma(\omega)$ and the d.c. resistivity
$\rho(T)$ within the extended $t$-$J$ model on a square lattice, as
relevant to high-$T_c$ cuprates, are reinvestigated using the
exact-diagonalization method for small systems, improved by performing
a twisted boundary condition averaging. The influence of the
next-nearest-neighbor hopping $t'$ is also considered. The behaviour
of results at intermediate doping is consistent with a
marginal-Fermi-liquid scenario and in the case of $t'=0$ for
$\omega>T$ follows the power law $\sigma \propto \omega^{-\nu}$ with
$\nu \sim 0.65$ consistent with experiments. At low doping $c_h<0.1$
for $T<J$ $\sigma(\omega)$ develops a shoulder at $\omega\sim
\omega^*$, consistent with the observed mid-infrared peak in experiments, 
accompanied by a shallow dip for $\omega < \omega^*$. This region is characterized by
the resistivity saturation, whereas a more coherent transport appears
at $T < T^*$ producing a more pronounced decrease in $\rho(T)$. The
behavior of the normalized resistivity $c_h \rho(T)$ is within a
factor of 2 quantitatively consistent with experiments in cuprates.
\end{abstract} 
 
\maketitle  

\section{Introduction}  

Since the discovery of high-$T_c$ cuprates, strong correlations among
electrons have been considered as the crucial reason for anomalous
transport properties in the normal state of these materials.
Prominent example is the d.c. resistivity obeying the well-known
linear law, $\rho \propto T$, in the intermediate (optimum) range of
hole doping accompanied by an anomalous but universal frequency
dependent optical conductivity $\sigma(\omega)$, phenomenologically
described via a marginal - Fermi - liquid (MFL) scenario \cite{varm}
or the quantum-critical behavior.\cite{mare} In the last decade the
emphasis has been centered on the experimental investigations of the
underdoped regime in various cuprates. The main signatures of this
regime are the kink in d.c. $\rho(T)$ at the crossover scale $T^*$
\cite{taka,ando1} and the appearance of a broad peak in the infrared
region (mid-IR) \cite{uchi} as well as of the pseudogap scale in
$\sigma(\omega)$.\cite{puch,timu} Recently, it has been established
within this regime that the mobility of holes increases with doping
concentration.\cite{ando}

From the theoretical point of view, it is not an easy task to construct an
analytical transport theory for strongly correlated
systems, starting from a microscopic model that describes the motion of
charge in a weakly or moderately doped antiferromagnetic (AFM)
insulator. Even at present such a theory is lacking. However,
numerical investigations of the prototype models, in particular of the
$t$-$J$ model, show that $\sigma(\omega)$ and $\rho(T)$ of cuprates
can be reasonably accounted for. Application of the finite-temperature
Lanczos method (FTLM) \cite{jplan,jprev} to small systems yields
results being most reliable in the regime of intermediate doping
\cite{jpsig,jpuni} where also the experimental window $T<1000$~K can
be probed best. In the latter regime, numerical results show an
overall agreement with the simple scaling $\sigma(\omega) \propto
(1-{\rm exp}(-\omega/T))/\omega$, with the only scale given by
temperature $T$.\cite{jpuni,jprev}

Recently, several aspects of the conductivity in cuprates have been
reopened urging for reconsideration and improvement of theoretical and
model results.  Detailed experimental studies in the optimally doped
cuprate Bi$_2$Sr$_2$Ca$_{0.92}$Y$_{0.08}$Cu$_2$O$_{8+\delta}$ (BSCCO)
\cite{mare} reveal for $\omega >T$ the $T$-independent power-law
behavior $\sigma(\omega) \propto \omega^{-\nu}$ with $\nu \sim 0.65$.
The authors attribute such a scaling to the vicinity of the
quantum-critical point.\cite{mare}

Another question is the existence and value of the resistivity
saturation in cuprates at low hole doping.\cite{cala,gunn} It has been
realized that cuprates at high $T$ and low doping show very large
resistivity $\rho$\cite{taka,ando} and thus violate the naive
Ioffe-Regel condition for metals, which implies a saturation of
$\rho(T)$ when the scattering length $L_s$ reaches the intercell
distance. On the basis of the $t$-$J$ model the modified saturation
value $\rho_{sat}$ has been derived \cite{cala} and its anomalous
large value has been ascribed to the kinetic energy being strongly
suppressed due to strong correlations, in particular $\langle
H_{kin}\rangle\propto c_h$ close to half-filling, where $c_h=N_h/N$ is
the hole number concentration. From the experimental view,
La$_{2-x}$Sr$_x$CuO$_4$ (LSCO) serves as a prototype cuprate with a
well controlled hole doping $c_h=x$ and stable material properties.
Previous and recent results \cite{taka,ando} in single-crystal LSCO
have revealed signatures of saturation at high $T>T^* \sim 500$~K.
Plausibly, the saturation phenomenon is related to the emergence of a
$T$ window, where $\sigma(\omega<\omega^*)$ is quite flat or even
develops a shallow dip. \cite{take,lee} The latter phenomenon is
clearly related to the appearance of a broad mid-IR peak at
$\omega \sim \omega^*$ first found at low doping in LSCO
\cite{uchi,take} and recently also in YBa$_2$Cu$_3$O$_y$
(YBCO). \cite{lee}

Recently, estimates were presented \cite{cala,gunn} that the $t$-$J$
model yields substantially smaller resistivity $\rho(T)$ than
experiments in LSCO for the regime of low doping. This apparently
leads to a conclusion that coupling to additional degrees of freedom,
in particular to phonons, might be essential to explain the large
resistivity.  This aspect is clearly of importance since it is related
to the central question, i.e., to what extent the $t$-$J$ model and
strong correlations alone can describe the physics of high-$T_c$
cuprates.

The existence of the mid-IR resonance in $\sigma(\omega)$ is
experimentally well established at low doping, \cite{uchi,take,lee}
however the consensus on its origin has not been reached yet. We note
that such a peak has been reproduced already in calculations of
$\sigma(\omega)$ for a single hole within the $t$-$J$ model at $T=0$
\cite{sega} and has been attributed to the string-picture of
incoherent hole motion leading to the shoulder at $\omega^* \sim
2J$. \cite{dago} However, a confirmation of this feature in more
realistic $T>0$ calculation was missing so far.

The aim of this paper is to give answers to above questions via a
systematic reinvestigation of $\sigma(\omega)$ and $\rho(T)$ within
the extended $t$-$J$ model, considering also the influence of the
next-nearest-neighbor (NNN) hopping $t'$ which has already been invoked in the
modeling of hole and electron doped cuprates\cite{tohy} to possibly
reveal the pronounced difference of $\sigma(\omega)$ between both classes
of cuprates. Additional hopping parameter $t'$ has been identified to
be important for the explanation of spectral properties, e.g., the
band dispersion as measured via the angle-resolved photoemission
experiments (ARPES).\cite{dama} Even more, $t'<0$ could be the
essential parameter discriminating various families of hole-doped
cuprates (LSCO, BSCCO etc.), \cite{raim,pava} in particular their
superconducting transition temperature $T_c$.  Relative to the
previous work \cite{jpsig,jprev} we are able to study somewhat larger
size systems using the FTLM. At the same time we improve the method by
introducing averaging over twisted boundary conditions (TBC),
described furtheron. The improvement shows up in a more controlled
behavior of $\sigma(\omega)$ at low $\omega$, being essential to
extract reproducible $\rho(T)$ at lower $T$. In the intermediate
doping we present calculations for systems up to $N=20$ sites to
clarify the universal behavior of $\sigma(\omega)$. In the low-doping
regime we study systems up to $N=26$ sites allowing us to establish
the emergence of a pseudogap scale in $\sigma(\omega)$ as well as the
onset of a more coherent transport for $T<T^*$. Our results still
remain restricted to $T>400$~K.  Nevertheless, they confirm the
qualitative behavior of $\sigma(\omega)$ in cuprates, with the
quantitative discrepancy in $\rho(T)$ compared to experiments within a
factor of 2 at most.

The paper is organized as follows. In Sec.~II the improved FTLM
method, employing the TBC averaging, is described. The comparison of
results with the usual fixed-boundary condition (BC) method is
presented. In Sec.~III the optimum doping results for $\sigma(\omega)$
and $\rho(T)$ are presented and discussed in connection with the
scaling behavior and experimental results for cuprates. In Sec.~IV the
low-doping regime is examined, with the emphasis on the emergence of
the shoulder corresponding to the mid-IR peak in $\sigma(\omega)$, the
resistivity saturation and the onset of a coherent transport for
$T<T^*$. Conclusions are given in Sec.~V.
 
\section{Numerical method}

In the following we study numerically the extended $t$-$J$ model
\begin{equation}
H=-\sum_{i,j,s}t_{ij} \tilde{c}^\dagger_{js}\tilde{c}_{is}
+J\sum_{\langle ij\rangle}({\bf S}_i\cdot {\bf S}_j-\frac{1}{4}
n_in_j), \label{tj} 
\end{equation}
on a square lattice, whereby we include besides the nearest neighbor
(NN) hopping $t_{ij}=t$ also NNN hopping $t_{ij}=t^\prime$.  Strong
correlations among electrons are incorporated via projected operators,
e.g., $\tilde{c}^\dagger_{is}= (1-n_{i,-s}) c^\dagger_{is}$, which do
not allow for a double occupancy of sites. The dependence of
$\sigma(\omega)$ on $t'$ has already been examined in connection with the
difference between the hole-doped and electron-doped cuprates.\cite{tohy}
However, the influence of $t'$ on the d.c. transport is
less evident and has not been studied systematically so
far.\cite{jprev} Plausibly, it is expected that $t'<0$, being
appropriate for hole-doped cuprates, leads to a larger frustration of
the AFM spin background and consequently to a larger resistivity. In
our study we will test the behavior at three different $t'=0$,
$t'=-0.15~t$ and $t'=-0.3~t$, respectively. We keep everywhere
$J=0.3~t$ as appropriate for cuprates, as well as we note that $t \sim
0.4$~eV for a direct comparison with experiments.

The main limitation to the validity of numerical results comes from
the finite-size effects which begin to dominate results at low
$T<T_{fs}$. As a criterion  for $T_{fs}$ we use the thermodynamic sum
\cite{jprev}
\begin{equation}
\bar{Z}(T)=\textrm{Tr}\exp[-(H-E_0)/T],
\label{thsum}
\end{equation}
calculated in a given system for fixed hole number $N_h$ and the
requirement $\bar{Z}(T_{fs})=Z^{\ast}\gg 1$. In the following we use
$Z^{\ast}\sim 30$.

Finite size effects can be substantially reduced by employing the TBC
averaging.\cite{poil,bonc} In a system with periodic BC
the latter is achieved by introducing an uniform vector
potential $\vec \theta$ modifying the hopping elements $t_{ij} \to
\tilde t_{ij} = t_{ij} ~\mathrm {exp}(i \vec \theta \cdot \vec
r_{ij})$ by a phase factor. We use furtheron $N_t$ different
phases $\vec \theta$ instead of an usually fixed BC $\theta=0$.
For an arbitrary operator $A$ we then perform the usual thermodynamic
averaging over $N_t$ different TBC
\begin{equation}
\left<A\right>=\frac{\sum_{j=1}^{N_t}\left<A\right>_jZ_j}{\sum_{j=1}^{N_t}Z_j}.
\end{equation}
where $\left<A\right>_j$ and $Z_j$ refer to the canonical FTLM
expectation values obtained for each fixed phase $\vec \theta$. In
order to preserve the translational invariance of the lattice
Hamiltonian in 2D we are allowed to study tilted square lattices with
$N$ sites where $N$ must be Pythagorean, i.e.,
$N=n^2+m^2$. Consequently the first Brillouin zone, where the phases
are chosen equidistantly, is a tilted square and $N_t$ is also chosen
Pythagorean, so that the phases form a regular square lattice as well.
Effectively, such a choice reproduces, e.g., for free fermions the
regular (square) lattice of $N N_t$ points in the ${\bf k}$ space
within the first Brillouin zone.

\begin{figure}[htb]  
\centering
\epsfig{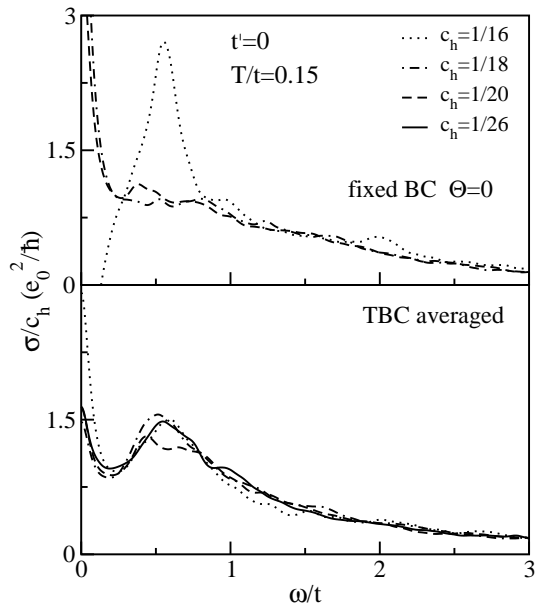}
\caption{Dynamical 2D mobility $\sigma/c_h$ for a single hole 
  $N_h=1$ in different systems $N=16,18,20,26$ evaluated at fixed
  $T/t=0.15$:
a) for fixed BC $\theta=0$, and b) with the TBC averaging.}
\label{fig1}
\end{figure}

There are several advantages of the TBC averaging procedure:

a) In the limit $T \to 0$ (for large enough $N_t$) this method
reproduces for an arbitrary quantity, e.g. for $\sigma(\omega)$, the
result corresponding to the ground state wavefunction $|\Psi_0(\vec{
\theta})\rangle$ with $E_0(\vec{\theta})=$~min. for a
chosen system with $N$ sites and $N_h$ holes.

b) In a finite (nonintegrable) system, even for $T>0$, we generally
expect the 2D optical conductivity of the form
\begin{equation}
\sigma(\omega)=2\pi D_c \delta(\omega)+ \sigma_{reg}(\omega),
\label{eqstiff}
\end{equation}
where
\begin{equation}
D_c=\frac{1}{2N}\langle \tau\rangle-\frac{1}{\pi e_0^2}
\int_{0^+}^{\infty}\sigma\left(\omega\right)d\omega,
\label{eqstiff1}
\end{equation}
$D_c$ is the charge stiffness representing the nondissipative part of
conductivity, emerging due to nonscattered coherent charge propagation
in a finite system with periodic BC, and $\langle\tau\rangle$ is the
kinetic stress tensor or the generalized kinetic energy which in the
case of NN hopping on a square lattice reduces to the usual kinetic
energy, i.e., $\langle \tau \rangle=- \left<H_{kin}\right>/2$.
\cite{mald} Since $D_c(T=0) \propto \partial^2 E_0(\vec{\theta})/
\partial^2 \vec{\theta} |_{E_0(\vec{\theta})=~min.}>0$,\cite{kohn} the
TBC averaging leads automatically to $D_c(T\to 0)>0$ as appropriate
for a metal. Note that previous calculations of $\sigma(\omega)$ at
fixed $\theta=0$ \cite{jpsig,jprev} generally suffered on a quite
system dependent $D_c(T\to 0)$ which was in many cases even negative.

c) In general, the TBC averaging reduces finite-size effects, although
it cannot completely eliminate them.  Results for $\sigma(\omega)$ are
thus more size and shape independent.  In Fig.~1 we present the
comparison of results for a single hole $N_h=1$ in different systems
$N=16,18,20,26$, as obtained for a fixed BC $\theta=0$ and when using
the TBC averaging. There is a clear difference in the low-$\omega$
regime where $D_c$ can be even negative for a fixed BC, while rather
consistent results are obtained when applying the improved method.
Moreover, the TBC averaged $\sigma(\omega)$ display quite evident and
consistent shoulder at $\omega^* \sim 0.5~t$, corresponding to the
mid-IR peak in cuprates, \cite{uchi,take,lee} while this feature is
hardly visible for fixed BC spectra.

In the following, we present results using typically $N_t\sim 10$ and
$M \sim 140$ Lanczos steps within each symmetry sector of the Hilbert
space. Since TBC averaging now also takes the role of random sampling we
do not perform any additional sampling over the initial Lanczos
wavefunction within each sector. In each of the available systems
$N=18,20,26$ we are able to reach $T_{fs}\sim 0.1~t$ for the intermediate
doping and somewhat higher for low doping. In the analysis of
$\sigma(\omega)$ spectra broadening $\epsilon=0.07~t$ is used.

\section{Intermediate - optimum doping}

One of the most striking facts in cuprates, recognized from the
beginning, is the universality of the charge response
$\sigma(\omega,T)$ in the vicinity of the optimum doping.  Whereas the
d.c. resistivity shows a linear variation $\rho(T) \propto T$, also
$\sigma(\omega)$ can be analyzed with an anomalous Drude-like
relaxation rate $1/\tau(\omega,T) \propto \omega + \xi T$.  The only
relevant $\omega$ scale seems to be given by $T$ itself, as summarized
within the phenomenological MFL scenario.\cite{varm} Recent study of
optimally doped BSCCO system has revealed, that at low $T$ in a broad range
of $T< \omega < 0.5$~eV one can describe results with a power law
$\sigma(\omega) \propto \omega^{-\nu}$, whereby $\nu \sim
0.65$.\cite{mare} Still, it is not possible to represent results for
$\sigma(\omega,T)$ in the whole $(\omega,T)$ range with a single
universal function of $\omega/T$ as, e.g., required within the
quantum-critical-point scenario.\cite{mare}

\vskip 0.8truecm
\begin{figure}[htb]
\centering
\epsfig{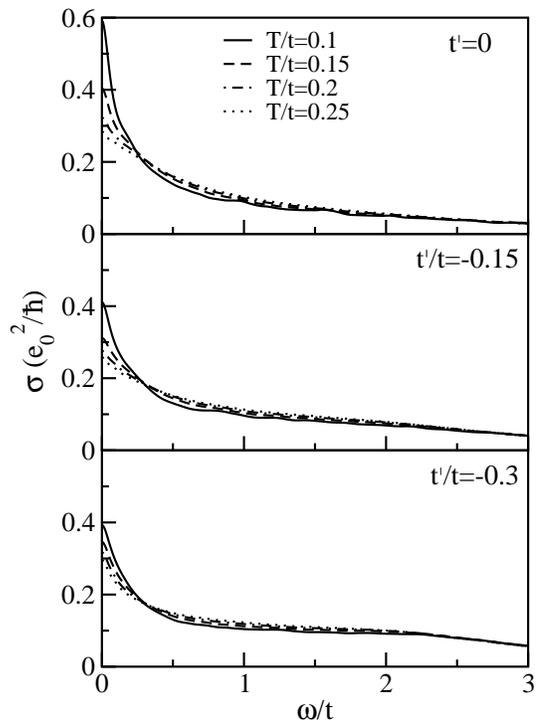}
\caption{2D optical conductivity $\sigma(\omega)$ for the intermediate doping $c_h=3/20$ 
as calculated for $N=20$ square lattice for different $T/t$ and 
$t'/t=0,-0.15,-0.3$.}
\label{fig2}
\end{figure}

Numerical results for $\sigma(\omega)$ within the $t$-$J$ model are
most reliable at intermediate doping, since there the lowest $T_{fs}$ can
be reached within a fixed system of $N$ sites.\cite{jprev} In Fig.~2
we present results for a lattice of $N=20$ sites and $N_h=3$ holes,
i.e., the hole doping $c_h=N_h/N=0.15$, at various $T/t$ and different
$t'/t=0,-0.15, -0.3$. Obtained 2D $\sigma(\omega)$ can serve as a test
for the scaling behavior. The optical sum rule,
\begin{equation}
\int_{-\infty}^\infty \sigma(\omega) d\omega= \frac{\pi e_0^2}{N}
\langle \tau \rangle,
\label{eqsum}
\end{equation}
requires a fast fall-off of $\sigma(\omega)$ for large
$\omega>\omega_c$, hence one can discuss the scaling only for 
$\omega<\omega_c \sim 2t$.

In Fig.~3 we study $\sigma(\omega)$ for $t'=0$ within the regime
$T<\omega <\omega_c$. We see that the fall-off is slower than $\propto
1/\omega$ and is well described with the power law $\sigma \propto
\omega^{-\nu}$ with $\nu <1$. Still, $\nu$ seems to depend slightly on
$t'$. The closest fit for $t'=0$ yields $\nu =0.65$ which is in
excellent agreement with experiments in BSCCO,\cite{mare} although for
the latter cuprates more appropriate model should be the one with $t'<0$.  For
$t'<0$ our results reveal a decreasing $\nu$. I.e., we get for
$t'/t=-0.15$ and $t'/t=-0.3$, $\nu=0.5$ and $\nu=0.42$, respectively.

\vskip 0.5truecm
\begin{figure}[htb] 
\centering
\epsfig{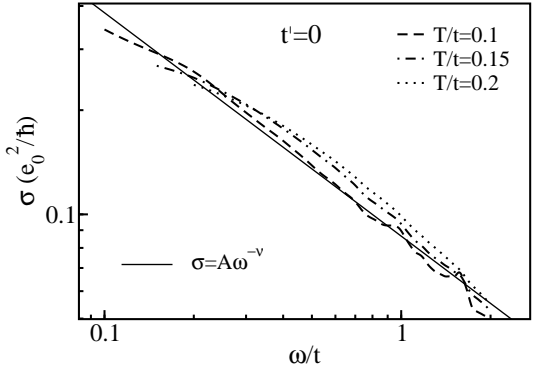}
\caption{Log-log plot of $\sigma$ vs. $\omega$  for $c_h=3/20$.
 Full line represents the closest fit to the power law
 with $\nu=0.65$ for $T/t=0.1$.}
\label{fig3}
\end{figure}

In order to come closer to the understanding of the scaling behavior, we
represent $\sigma(\omega)$ in a general form following the Kubo
formula,
\begin{equation}
\sigma(\omega) =  C(\omega) \frac{1 - e^{-\omega/T}}{\omega}  ,
\label{eqsig}
\end{equation}
where $C(\omega)$ is current-current correlation function
\begin{equation}
C(\omega) = \text{Re} \int_0^{\infty} dt~ e^{i \omega
t }\langle j(t) j \rangle. \label{eq3}
\end{equation}
It has been observed that the scaling behavior of $\sigma(\omega,T)$
within the $t$-$J$ model at intermediate doping can be well described
up to the cutoff $\omega <\omega_c$ by a parameter-free form of
Eq.~(\ref{eqsig}) with $C(\omega) \sim C_0$ where $C_0$ is
$T$-independent. Such a form is clearly a restricted version of the
MFL behavior \cite{varm} and reproduces both $\rho \propto T$ law and
$\sigma(\omega \gg T) \propto 1/\omega$.  In view of present improved
results presented in Fig.~3 we can analyze the deviation from the
simple $C(\omega)=C_0$ form. In Fig.~4 we show $C(\omega)$ for $t'=0$
and various $T/t$. In this case the spectra are broadened with the
factor $\epsilon=0.1t$. The main characteristic is that $C(\omega)$ is
an increasing function for $\omega<\omega_c$, consistent with
effective $\nu<1$.
    
The origin of a very broad range of validity of $\sigma \propto
\omega^{-\nu}$ has not become clear yet. Note that $\nu<1$ requires an
increasing $C(\omega)$ which in normal metals is supposed to have a
Lorentzian form with a characteristic width determined by a Drude
relaxation rate $1/\tau<T$. The interpretation of the anomalous
(constant or even increasing) $C(\omega)$ has been proposed by one of
the authors.\cite{prel} By performing a simple decoupling of
$C(\omega)$ in terms of the single-electron spectral functions $A({\bf
  k}, \omega)$ and neglecting the vertex corrections, one gets
\begin{eqnarray}
&&C(\omega)=\frac{2\pi e_0^2}{N} \sum_{\bf k} (v_{\bf k}^\alpha)^2 
\int {d\omega^\prime}  \times \nonumber \\
&&f(-\omega^\prime) f(\omega^\prime-\omega)
A({\bf k},\omega^\prime) A({\bf k},\omega^\prime-\omega),
\label{eqcan}
\end{eqnarray}
where $f$ is the Fermi function and $v_{\bf k}^\alpha$ unrenormalized
band velocities. We represent $A({\bf k}, \omega)$ for quasiparticles
close to the Fermi energy as
\begin{equation}
A({\bf k}, \omega)=\frac{1}{\pi}\frac{Z_{\bf k}\Gamma_{\bf k}}
{(\omega-\epsilon_{\bf k})^2+\Gamma^2_{\bf k}},
\label{eqak}
\end{equation}
where parameters $Z_{\bf k},\Gamma_{\bf k},\epsilon_{\bf k}$ in
general dependent on $T$. In order to reproduce the MFL scaling of
$\sigma(\omega)$ we have to assume the MFL form for the damping, i.e.,
$\Gamma = \gamma (|\omega| + \xi T)$ as well as neglect the ${\bf k}$
dependence of $\Gamma$ and $Z$. With these simplifications we arrive
at
\begin{equation}
C(\omega)=\bar C \int d\omega' f(-\omega')f(\omega'-\omega)
\frac {\bar \Gamma(\omega,\omega')}
{\omega^2+ \bar \Gamma(\omega,\omega')^2}, \label{eqc}
\end{equation}
where $\bar \Gamma(\omega,\omega')=\Gamma(\omega')
+\Gamma(\omega'-\omega)$. It has been noted \cite{prel} that
$C(\omega) \sim C_0$ appears already for $\gamma \sim 0.3$, while for
$\gamma>0.3$ it is an increasing function of $\omega/T$. We
should also remind that $\gamma>0.3$ is consistent with ARPES
experiments in BSCCO, where along the nodal direction \cite{vall} as
well as in other directions \cite{kami} $\gamma \sim 0.7$ was
obtained.

\vskip 0.5truecm
\begin{figure}[htb]
\centering
\epsfig{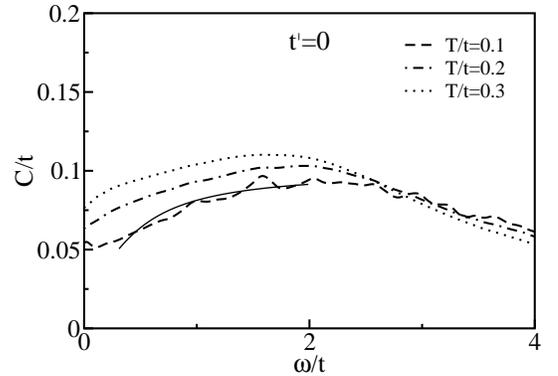}
\caption{$C(\omega)$ for intermediate doping $c_h=3/20$ 
  and various $T/t$. Full line represents the analytical result from
  Eq.~(\ref{eqc}) with $\gamma=1.2$, $\xi=\pi$ and $T/t=0.1$.}
\label{fig4} 
\end{figure}

For comparison with the numerical result at the lowest $T/t=0.1$, we
display in Fig.~4 $C(\omega)$ following from Eq.~(\ref{eqc}) with
$\gamma =1.2$ and $\xi=\pi$. The qualitative behavior is satisfying,
with a visible deviation only at low $\omega<T$. This indicates that
anomalous $C(\omega)$ as well as the power law $\sigma(\omega) \propto
\omega^{-\nu}$ with $\nu<1$ can be qualitatively described with the
assumption
that the quasiparticle damping is of the MFL-form with large
$\gamma >0.3$. Still, the description with the simple Eq.~(\ref{eqc}) cannot
quantitatively account for the whole region $(\omega,T)<\omega_c$.

\vskip 0.5truecm
\begin{figure}[htb]
\centering 
\epsfig{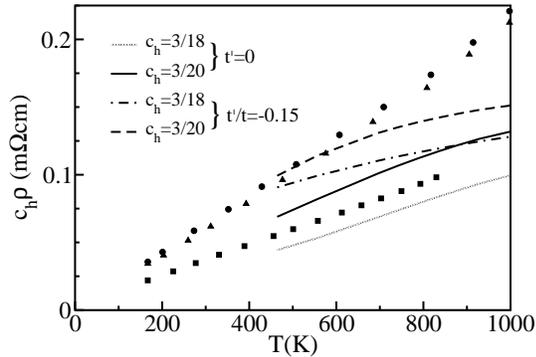}
\caption{Normalized resistivity $c_h\rho(T)$ at intermediate doping 
for $t'=0$ and $t'/t=-0.15$,
compared with the experimental data for LSCO. Squares denote results for
$x=0.15$,\cite{taka} while circles and triangles are for $x=0.15,
0.18$,\cite{ando} respectively.}
\label{fig5} 
\end{figure}

Let us discuss the quantitative comparison of our novel results for
$\rho(T)=c/\sigma(\omega=0,T)$ (we use $c=6.6\textrm{\AA}$) with the
experiments at intermediate doping. In Fig.~5 we compare experimental
results for the normalized resistivity $c_h\rho$ of LSCO for $x=0.15,
0.18$ \cite{taka,ando} with corresponding FTLM results for $N_h=3$ in
systems with $N=18,20$ sites in the case of $t'=0$ and $t'/t=-0.15$,
respectively. Results for $t'=0$ are well consistent with the linear
$\rho \propto T$ law in the $T$ regime of experimental relevance. It
is characteristic that $t'/t=-0.15$ results reveal somewhat larger
$\rho$ at lower $T$ while also the presumed linearity is reached at
lower $T<1000$~K. Taking into account the variation of different
experimental data the quantitative agreement is satisfying although we
allow for the possibility that our model results underestimate $\rho$.

\section{Low-doping regime}

In previous numerical studies of $\sigma(\omega,T)$ results for the
low-doping regime \cite{jpsig,jprev} were less conclusive. The reasons
were the following: a) $T_{fs}$ increases as $c_h\to 0$. So even
$T_{fs} \sim 0.15~t$ was unreachable in the systems available with the
FTLM a decade ago. b) At the lowest doping, i.e., for the case of a
single hole $N_h=1$, there is a significant nondissipative
contribution $D_c \neq 0$ at $T \sim T_{fs}$.  This indicates that the
effective scattering length $L_s$ might become larger than the system
size $\sqrt{N}$.  In addition, $D_c$ is varying quite uncontrollably
between different system sizes when the FTLM is performed with a fixed
BC. In fact, as shown in Fig.~1 in several cases one gets unphysical
$D_c<0$.  c) Systematic experimental studies of $\rho(T)$ and
$\sigma(\omega)$ in cuprates at low doping have become available
more recently.\cite{ando,take,lee}

\begin{figure}[htb]
\centering
\epsfig{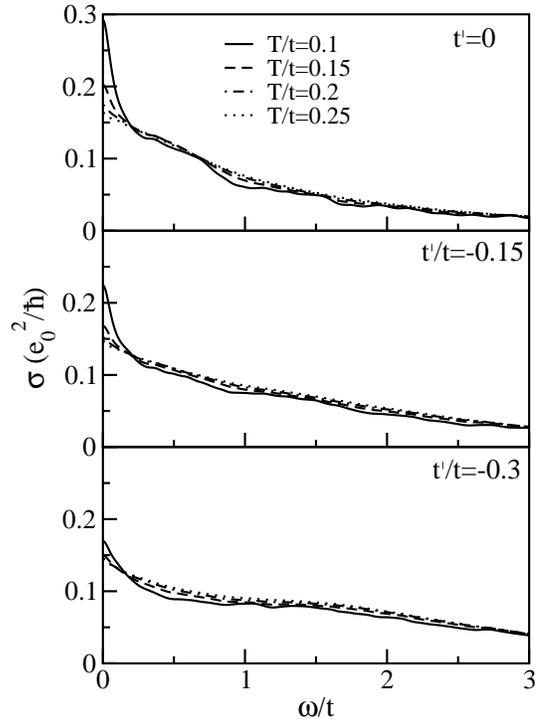}
\caption{2D $\sigma(\omega)$ for underdoped $c_h=2/20$, at different $T/t$ and 
$t'/t$.}
\label{fig6}  
\end{figure}

The TBC averaging substantially improves the FTLM results for
$\sigma(\omega)$, as seen in Fig.~1.  They are much less system-size
dependent even in the presence of a finite $D_c>0$. First the results
for planar $\sigma(\omega)$ in the underdoped case are presented in
Fig.~6.  We note that the behavior is generally very similar to the
one at intermediate doping in Fig.~2. Still, there appears to be
already some build-up of a shoulder at, e.g., $\omega \sim 0.5~t$ for
$t'=0$, as a precursor of the mid-IR scale, hence $\sigma(\omega)$
cannot be well described by a power law $\sigma \propto \omega^{-\nu}$
any more. Results for $c_h=0.1$ can be considered as a crossover to
substantially different behavior in the low-doping regime which we
discuss furtheron.

\begin{figure}[htb]
\centering
\epsfig{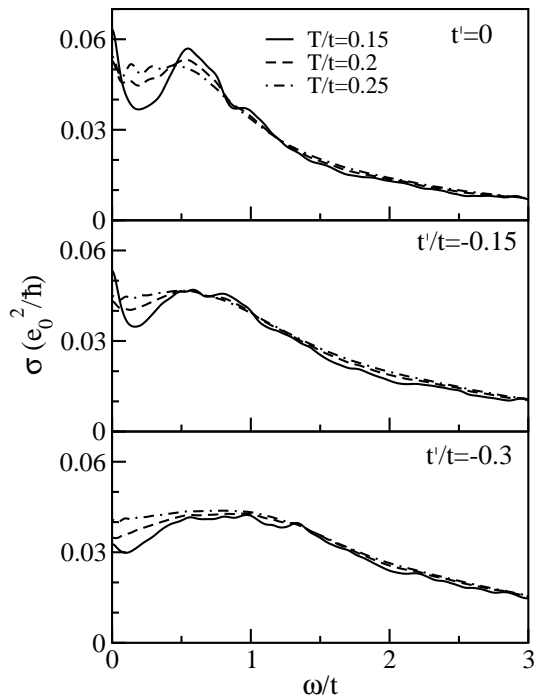}
\caption{2D $\sigma(\omega)$ for low doping,
calculated for $c_h=1/26$ at different $T/t$ and $t'/t$.}
\label{fig7}
\end{figure}

In Fig.~7 we present results for $\sigma(\omega)$, as obtained for the
largest system with $N=26$ sites and a single hole $N_h=1$. Very
similar results were obtained also for $N_h=1$ in systems with $N=18,
20$ sites. The behavior of $\sigma(\omega)$ is clearly different from
the one belonging to the optimum doping in Fig.~2. The main difference
comes from the emergence of a shoulder at $\omega^*$ which further
depends on $t'$.  While $\omega^* \sim 0.5~t$ at $t'=0$, it increases
to $\omega^* \sim t$ for $t'=-0.3 t$.  The shoulder starts to build up
in $\sigma(\omega)$ for $T<J$ which is a quite clear indication that
the observed effect is related to the onset of short-range AFM
correlations, confirming the relevance of the string-picture of the
incoherent hole motion.\cite{dago} Lowering $T<J$ we are in a regime
of nearly constant $\sigma(\omega<\omega^*)$ which at the same time
induces the flattening or the saturation of $\rho(T)$. It should be
noted that our shoulder can plausibly be related to the mid-IR
resonance in cuprates,\cite{uchi,take,lee} in fact the position of
both is even quantitatively well in agreement.
At the same time a dip in $\sigma(\omega)$ has also been
observed in recent experiments in LSCO at low doping $x=0.08$ and
above the crossover temperature $T>T^* \sim 500$~K,\cite{take} as well
as in YBCO. \cite{lee}

At decreasing $T<T^*<J$ our results indicate the onset of a
nondissipative contribution $D_c(T)>0$.  Calculated $D_c(T)$ are
surprisingly consistent for different systems $N=18, 20, 26$, as
presented in Fig.~8. Plausibly one expects that in a large enough
system additional scattering channels would lead to the broadening of
the $\omega =0$ peak. Nevertheless its weight should not change
considerably and also its width should remain at least narrower than
the pseudogap scale $\omega^*$.  Quite abrupt onset of $D_c>0$ at $T^*
\sim 0.15~t \sim 600$~K is quite consistent with the experimentally
observed scale $T^*$ at low doping, identified as the kink in
$\rho(T)$.\cite{taka,ando} To evaluate proper $\rho(T<T^*)$ we are
missing the scattering mechanism within our small systems. For
convenience we evaluate $\rho(T<T^*)$ by using spectra broadening
$\epsilon$ as elsewhere in our analysis.

\vskip 0.3truecm
\begin{figure}[htb]
\centering
\epsfig{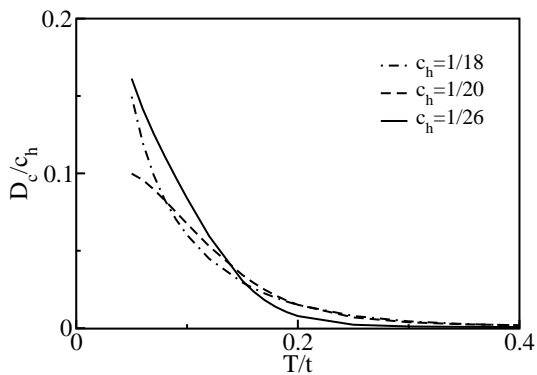}
\caption{Normalized charge stiffness $D_c/c_h$ as calculated for $N_h=1$ and
$N=18,20,26$ in the case of $t'=0$.}
\label{fig8} 
\end{figure}

Let us continue with the results for the normalized 2D resistivity
$c_h \rho$ presented in Fig.~9 in a large range of temperatures $T<t$
and for a wider regime of doping up to the 'optimum' doping
$c_h=0.15$. For large $T>0.5~t$ all curves merge on the 'trivial'
linear dependence $d\rho /dT=\zeta\rho_0 k_B/c_ht$ with $\zeta \sim
0.4$, as follows, e.g., from the high-$T$ expansion for
$\rho(T)$.\cite{jpsig,jprev} For the intermediate doping there appears
a steady crossover at $T\sim 0.25~t$ to the low-$T$ linear law, as
discussed in Sec.~III. On the other hand, low-doping results reveal
quite evidently a pseudo-saturation of $\rho(T)$ within the $T$
window $T^*<T<0.5~t$, closely related to the appearance of a flat and
$T$-independent $\sigma(\omega)$ in this regime. We can compare our 2D
saturation value $\rho_{sat}$ with the estimate in Refs.\cite{cala,gunn}, obtained
by assuming $\sigma(\omega<W) \sim $~const. with $W$ being a typical
band-width. Since $\langle H_{kin}\rangle /N \sim -3.4tc_h(1-c_h)$ for
2D $t$-$J$ model \cite{cala} one gets from Eq.~(\ref{eqsum}) at low
doping $c_h \to 0$,
\begin{equation}
c_h \rho_{sat} \sim 0.37\frac{W}{t}\rho_0 \sim 0.75\rho_0,
\label{eqsat}
\end{equation}
assuming an effective width of $\sigma(\omega)$ spectra $W/t \sim 2$
as estimated from Fig.~7, and $\rho_0=\hbar/e_0^2$ denotes the universal sheet
resistivity. Our saturation value in Fig.~9 is
well in agreement with the one following from Eq.~(\ref{eqsat}).

\vskip 0.5truecm
\begin{figure}[htb]
\centering
\epsfig{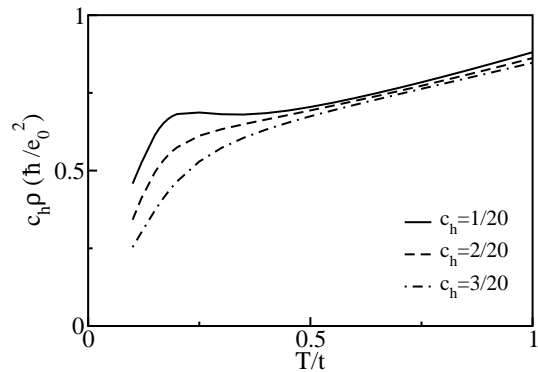}
\caption{Normalized 2D resistivity  $c_h \rho$ vs. $T/t$ for
  different $c_h$ and $t'=0$.}
\label{fig9}  
\end{figure}

Finally, we compare in Fig.~10 our results for $c_h \rho(T)$ at $t'=0$
with the experimental values for LSCO at low doping.  \cite{taka,ando}
First we note that the calculated values for $c_h \rho(T)$ in the low doping regime are
larger compared to those at intermediate doping
in Fig.~5 as observed in experiments. In the same way, it follows from Fig.~10 that $c_h
\rho(T)$ is larger for $c_h=0.05$ than for $c_h=0.1$. Comparing our
results for the lowest doping $c_h=1/20, 1/26$ we can observe the behavior
with a kink at $T \sim T^*$ similar to experiments.  The values
themselves are somewhat lower than those obtained by Ando et al.
\cite{ando}, and within a factor of $2$ to previous data.\cite{taka}

\vskip 0.5truecm
\begin{figure}[htb]
\centering
\epsfig{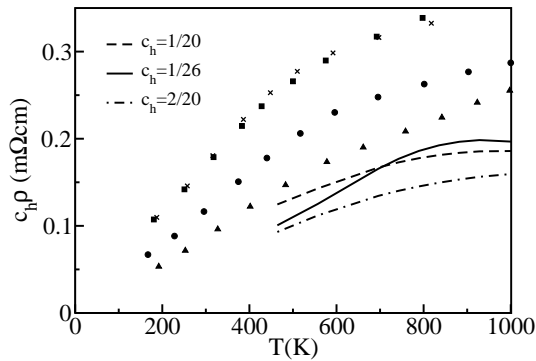}
\caption{Normalized resistivity $c_h \rho(T)$ at low doping for
$t'=0$ compared to experimental results for LSCO: $x= 0.04,
0.07$,
crosses and squares,\cite{taka} and $x= 0.04, 0.08$, circles and
triangles,\cite{ando} respectively.}
\label{fig10} 
\end{figure}

\section{Conclusions}

Let us summarize and discuss our results for the optical conductivity
$\sigma(\omega)$ and the d.c. resistivity $\rho(T)$ within the
extended $t$-$J$ model. The TBC averaging method, used in our studies,
provides a clear improvement of the FTLM exact-diagonalization
studies of small systems. In particular, the advantage is evident in
the low-doping regime, which is one of the central points in our
analysis.

Our study confirms that the transport properties substantially differ
between the low-doping regime $c_h <0.1$ and the intermediate doping,
in our case $c_h \sim 0.15$. With respect to the NNN hopping $t'$, our
results indicate that the transport quantities are not much sensitive
to its value, at least for $T>T_{fs}$ which we are able to study
numerically. General trend is that $t'<0$, as appropriate for
hole-doped cuprates, slightly increases $\rho(T)$ as well as the
shoulder (mid-IR) scale at low doping. At the same time, the
characteristic behavior at intermediate doping moves with $t'<0$
towards the properties at lower doping. In particular, properties at
$t'=0$ and $t'/t=-0.15$ are qualitatively similar, while $t'/t=-0.3$
already exhibits more pronounced deviation even at intermediate
doping.

In this paper we present results for $J/t=0.3$ only, as believed to be
the best parameter choice for cuprates. Still, additional results
obtained for $J/t=0.4$ (not shown here) reveal that both
$\sigma(\omega)$ and $\rho(T)$ do not change appreciably as far as we
are dealing with the strong correlation regime $J<t$.

At intermediate doping $\sigma(\omega)$ is well consistent with the
anomalous scaling described within the MFL scenario. The overall
behavior is close to the universal form, Eq.~(\ref{eqsig}), with
$C(\omega)=C_0$ in a wide range $\omega<\omega_c\sim 2~t$ and
$T<J$. On the other hand, we can establish also deviations from the
above simple form, in particular $C(\omega)$ is an increasing function
of $\omega$. At $\omega>T$ the scaling is thus consistent with the
power law $\sigma \propto \omega^{-\nu}$ with $\nu<1$, in particular
$\nu \sim 0.65$ for $t'=0$ very close to recent experimental result
for BSCCO,\cite{mare} although one would expect better agreement with
$t'<0$ results in the latter cuprates.

The anomalous behavior of $\sigma(\omega)$ and $C(\omega)$ can be
partly understood in terms of a simple representation of electron
spectral functions, Eq.~(\ref{eqc}), where the quasiparticle damping
$\Gamma = \gamma (|\omega| + \xi T)$ is of the MFL form with
large $\gamma>0.3$. It should be also noted that $C(\omega)$,
Eq.~(\ref{eqc}), leads in general to an analytical dependence at small
$\omega$, i.e., $T \sigma(\omega \to 0) \sim B + A
(\omega/T)^2$,\cite{prel} in contrast to the simple $C(\omega)=C_0$
assumption. Still, we find that the same input for $\gamma$ which
can describe the high-$\omega$ behavior cannot quantitatively account
for experimentally observed $\omega/T \to 0$.\cite{mare}

The origin of the anomalous MFL-like behavior of $\sigma(\omega)$
has not been clarified yet. The scaling range $\omega<2~t$
is surprisingly broad, since one would expect that it might be also
determined by $J$. On the other hand, the behavior qualitatively
changes for $T>J$. \cite{jpuni,jprev} This is another sign, that we
are dealing with an unusual frustrated fermionic system and possibly
not with the quantum phase transition \cite{mare} which should possess
only a single scale in the $(\omega,T)$ diagram.

At low doping $c_h<0.1$ the behavior changes qualitatively.
$\sigma(\omega)$ exhibits a shoulder at $\omega^*$ and flat region or
a weak dip for $\omega<\omega^*$, as found also in recent measurements
in LSCO \cite{take} and YBCO. \cite{lee} The shoulder is clearly
related to the mid-IR peak found in experiments, whereby we find
$\omega^* > 2J$ being in a reasonable quantitative agreement. The
shoulder sets in for $T<J$ and is plausibly related to the onset of
short-range AFM correlations hence this confirms the magnetic origin
of the mid-IR resonance and its spin-string interpretation.\cite{sega,dago}
Another manifestation of the same phenomenon is the
saturation of $\rho(T)$ in the temperature window $T^*<T<J$ which is
in agreement with experiments in cuprates.\cite{gunn}

For $T<T^*$ we observe the onset of a coherent transport in our small
systems, which shows up in a rather abrupt increase of the charge
stiffness $D_c(T)$. The appearance of $D_c(T)>0$ means that the
scattering length $L_s(T)$ becomes larger than the system size
$\sqrt{N}$. In this connection, it is quite puzzling to explain why
$L_s(T)$ is larger in the low doping regime than at intermediate
doping.  I.e., we find $D_c \sim 0$ down to $T \sim T_{fs}$ at
$c_h=0.15$, whereas evidently we get $D_c>0$ at the same $T$ in the
low-doping regime. This apparent contradiction is, however, consistent
with recent experimental confirmation of a quite coherent transport at
low doping for $T<T^*$ and the concept of a nodal metal. \cite{ando,lee}

While qualitative behavior of $\rho(T)$ is well in agreement with
experiments, let us finally comment on the quantitative comparison.
At intermediate doping the normalized resistivity $c_h \rho(T)$ lies tolerably
within the range of existing experimental values for LSCO, taking into
account also some variation between different published results. At
low doping our results seem to underestimate somewhat $\rho(T)$.
Still, the discrepancy (depending again on different experimental data) is
at most of a factor of $2$. Whether this rather modest disagreement is an
indication of the possible relevant role of degrees of freedom outside the
$t$-$J$ model, remains an open question.

We acknowledge the discussion with O. Gunnarsson, and the use of
unpublished data by Y. Ando and coworkers. This work was funded by the
Ministry of Higher Education, Science and Technology of Slovenia under the grant
Pl-0044. One author (P.P.) also acknowledges the support of the
Alexander von Humboldt Foundation.

\end{document}